# A Two-Stage Deep Learning Framework for Segmentation of Ten Gastrointestinal Organs from Coronal MR Enterography

Ashiqur Rahman[1], Md. Abu Sayed[1], Md. Sharjis Ibne Wadud[1], Md. Abu Asad Al-Hafiz[2], Adam Mushtak[3], and Muhammad E. H. Chowdhury[4,*]

[1]Department of Biomedical Physics and Technology, University of Dhaka, Dhaka 1000, Bangladesh. Email: ashiq@bmpt.du.ac.bd (AR), sayed@bmpt.du.ac.bd (MAS), sharjis@du.ac.bd (MSIW).

[2]Department of Biomedical Engineering, Jashore University of Science and Technology, Jashore 7408, Bangladesh. Email: asadalhafiz01@gmail.com (MAAAH).

[3]Department of Radiology, Hamad Medical Corporation, Doha, Qatar. amushtak@hamad.qa (AM).

[4]Department of Electrical Engineering, Qatar University, Doha 2713, Qatar. Email: mchowdhury@qu.edu.qa (MEHC).

Corresponding author: Muhammad E. H. Chowdhury (mchowdhury@qu.edu.qa).

# Abstract

**Objective:** Accurate segmentation of gastrointestinal (GI) organs in magnetic resonance enterography (MRE) is critical for diagnosing inflammatory bowel disease (IBD. However, anatomical variability, class imbalance, and low tissue contrast hinders reliable automation. This study proposes a dual-stage deep learning framework for organ-specific segmentation of GI structures from coronal MRE images to address these challenges.

**Methods:** A publicly available MRE dataset of 3,195 coronal T2-weighted HASTE slices from 114 IBD patients was used. Initially, a DenseNet201-UNet++ model generated coarse masks for ROI extraction. DenseNet121–SelfONN–UNet was then trained on organ-specific patches. Extensive data augmentation, normalization, five-fold cross-validation, and class-specific weighting were applied to mitigate severe class imbalance, particularly for the appendix.

**Results:** The initial stage achieved strong organ localization but underperformed for the appendix; class weighting improved its DSC from 6.76% to 85.76%. The second-stage DenseNet121–SelfONN U-Net significantly enhanced segmentation across all GI structures, with notable DSC gains (cecum +23.62%, sigmoid +18.57%, rectum +17.99%, small intestine +16.06%). Overall, the framework achieved mDSC 88.99%, mIoU 84.76%, and mHD95 6.94 mm, outperforming all baselines.

**Conclusion:** This framework demonstrates the effectiveness of a coarse-to-fine, organ-aware segmentation strategy for intestinal MRE. Despite higher computational cost, it shows strong potential for clinical translation and enables anatomically informed diagnostic tools in gastroenterology.

1. **Introduction**

The gastrointestinal (GI) organs play a critical role in digestion, nutrient absorption, and waste elimination, while also serving as a central component of the body's neuro-immune system [1], [2]. Beyond its physiological functions, the GI tract is linked to overall health through its interaction with the enteric nervous system and its communication with the central nervous system [3], [4]. Anatomically, it comprises distinct organs including the stomach, duodenum, small intestine, appendix, cecum, and various sections of the colon and rectum where each has specialized digestive functions. Maintaining the structural and functional integrity of these organs is essential for systemic health and homeostasis [5]. An anatomical structure of human digestive system is shown in Fig. 1.

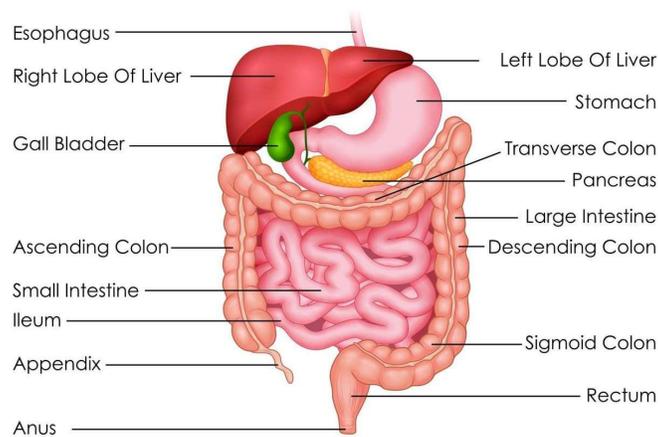

Fig. 1. Overview of the anatomy of human digestive system [6].

Inflammatory Bowel Disease (IBD), which comprises Crohn's disease (CD) and ulcerative colitis (UC), is a growing global concern due to its impact on the GI tract. Epidemiological data indicate an incidence rate of 10 to 150 cases per 100,000 individuals [7]. The high incidence rate necessitates accurate diagnosis. Magnetic Resonance Enterography (MRE) has emerged as a preferred modality for diagnosing IBD in GI organs due to its excellent soft-tissue contrast, radiation-free imaging, and multiplanar capabilities. It is especially useful for visualizing deeper regions of the bowel which are inaccessible via endoscopy [10], [11], [12]. MRE's diagnostic potential makes it useful for the automated segmentation of GI organs. Accurate segmentation of GI organs plays a vital role in early disease detection, improving diagnostic precision, and supporting treatment planning for conditions such as IBD [13]. In radiation therapy, it enables precise targeting of abdominal organs while minimizing damage to surrounding healthy tissues [14], [15]. Despite its importance, manual segmentation is both time-consuming and prone to inter-observer variability. Another challenge lies in the manual segmentation of specific intestinal regions such as the ascending colon, descending colon, transverse colon, and sigmoid colon. These sections are difficult to delineate due to overlapping anatomical boundaries, morphological variability, and inter-patient differences. Furthermore, the appendix remains largely underrepresented due to its small size, inconsistent anatomical location, and subtle radiological presentation contribute to its frequent exclusion from automated approaches, despite its clinical significance. These challenges have fueled growing interest in deep learning-based methods, which offer rapid, accurate, and highly consistent segmentations, making them indispensable for reliable clinical decision-making. Although notable advancements have been made in gastrointestinal organ segmentation using deep learning, several significant limitations continue to hinder clinical implementation. The most existing models are trained on limited datasets, which undermines generalizability across diverse patient populations [20], [21].

To address these limitations, this study proposes a comprehensive two-stage deep learning framework for the automated segmentation of ten anatomically distinct gastrointestinal organs including the stomach and nine bowel components from MRE images. The primary objective of this work is to develop a modular, anatomically aware, and clinically viable segmentation pipeline capable of handling the challenges associated with multi-organ abdominal MRE analysis. This includes addressing class imbalance, preserving spatial continuity, and enhancing segmentation accuracy in low-contrast or ambiguous anatomical regions. The key contributions of this study are as follows:

- We propose a novel dual-stage segmentation framework that combines coarse global localization with refined organ-specific segmentation, designed explicitly for MRE scans.
- We introduce a novel deep learning architecture in the second stage, featuring a pretrained DenseNet121 encoder and a SelfONN-based decoder, to enable nonlinear, task-adaptive feature learning.
- Proposed a strategy tackle severe class imbalance while preserving segmentation accuracy across the remaining organ classes.
- We demonstrate effective and anatomically consistent segmentation across ten bowel and GI structures, including clinically underrepresented organs such as the appendix, cecum, and sigmoid colon.

This paper is structured as follows: Section 2 reviews prior work on gastrointestinal organ segmentation, highlighting recent advancements and remaining challenges. Section 3 outlines the methodology, including dataset details, preprocessing steps, and the proposed two-stage segmentation framework. Section 4 presents experimental results, comparing our approach with baseline models and class-wise performance improvements. Finally, Section 5 summarizes the findings and discusses future directions for advancing abdominal MRE segmentation.

## 2. Literature Review

Segmenting GI organs from medical imaging remains a persistent challenge due to their anatomical complexity, variability in shape and spatial location, and the inherent difficulty in distinguishing organ boundaries especially in MRE scans. These challenges are further compounded by MRE-specific issues such as motion artifacts, non-uniform bowel distension, and subtle contrast differences between adjacent soft tissues. Historically, research in GI segmentation has focused on isolated organs or regions, limiting clinical applicability in comprehensive diagnostic workflows.

For instance, Gonzalez et al. [22] introduced a 2.5D deep learning framework for segmenting the sigmoid colon in CT images, integrating 2D and 3D convolutions for refinement. While this method achieved a DSC of 0.88 with prior anatomical localization, it was constrained to a single GI organ and did not address broader inter-organ variability or scalability to multiple structures. Lamash et al. [23] applied a 3D residual U-Net with curved planar reformatting to pediatric MRI, segmenting the small bowel into lumen, wall, and background. However, this model was specialized for a specific organ and patient population, limiting its relevance to general-purpose GI segmentation, especially in adults.

Van Harten et al. [24] expanded on temporal dynamics by segmenting the small intestine from 3D cine-MRI using a multi-task convolutional neural network and stochastic tracking. Despite reporting strong performance (DSC of 0.88 in healthy and 0.79 in IBD patients), the model remained limited to a single GI organ and was not designed to scale to broader anatomical coverage. Dellschaft et al. [25] analyzed small bowel function in cystic fibrosis patients using cine-MRI and Haralick texture features, but this work did not incorporate explicit segmentation, rendering it unsuitable for structural delineation or clinical quantification.

Orellana et al. [26] developed a probabilistic modeling framework using dual-phase registration to segment four colon regions such as ascending, transverse, descending, and sigmoid from T1-weighted fat-suppressed MRI. Although this marked a step toward multi-organ segmentation, the model still excluded key GI organs such as the duodenum, ileum, cecum, and appendix. Additionally, the reliance on mesh-based shape priors restricts its adaptability to organs with high inter-subject morphological variation. Ding et al. [27] proposed a 3D ResUNet with active contour refinement to segment six abdominal organs on T2-weighted MRI, including portions of the bowel. While this hybrid model improved spatial precision, its handcrafted post-processing steps reduced scalability and reproducibility. More importantly, the segmentation scope excluded smaller GI organs, such as the appendix and duodenum—and was not tailored for MRE, a modality that presents distinct image characteristics and motion dynamics.

Despite these developments, the studies remain constrained in several critical ways. Brem et al. [28] reviewed the role of deep learning in MRE for CD diagnosis and identified key challenges, including limited sample sizes, heterogeneous methodologies, and a lack of external validation. Most methods are tailored for segmenting between one and six abdominal organs, with limited exploration of comprehensive GI tract coverage. The challenge of class imbalance particularly in identifying small or underrepresented structures such as the appendix or sigmoid colon remains unaddressed. Furthermore, few studies utilize MRE as the primary imaging modality, despite its clinical importance in assessing inflammatory bowel diseases. The lack of MRE-focused models limits applicability in settings where soft-tissue contrast and motion artifacts present distinct challenges.

To address these gaps, this study introduces a novel dual-stage deep learning framework specifically designed for MRE-based segmentation of ten gastrointestinal structures. It begins with a model employed for initial coarse segmentation, followed by class-wise binary refinement networks. This modular, organ-aware approach, enhanced by nonlinear SelfONN transformations, enables accurate localization across diverse organ morphologies and offers a scalable, clinically meaningful improvement over existing methods.

3. Materials and Methods

This study proposes a dual-stage segmentation framework for the pixel-wise delineation of ten distinct GI organs from MRE scans. As illustrated in Fig. 2, the workflow begins with pre-processing, where annotated Neuroimaging Informatics Technology Initiative (NIfTI) images are converted to 2D Portable Network Graphics (PNG) slices. The dataset is split, augmented, and normalized, with 5-fold cross-validation and class reweighting applied to address imbalance. In the first stage, a multiclass DenseNet201 with UNet++ network was trained to generate coarse masks. These masks are then used to extract class-specific regions of interest (ROIs) for each of the ten GI organs. In the second stage, individual class-wise models are trained on these ROIs to refine segmentation results, enabling enhanced accuracy for underrepresented organs. This modular design enables both global localization and local refinement across diverse GI organ segments.

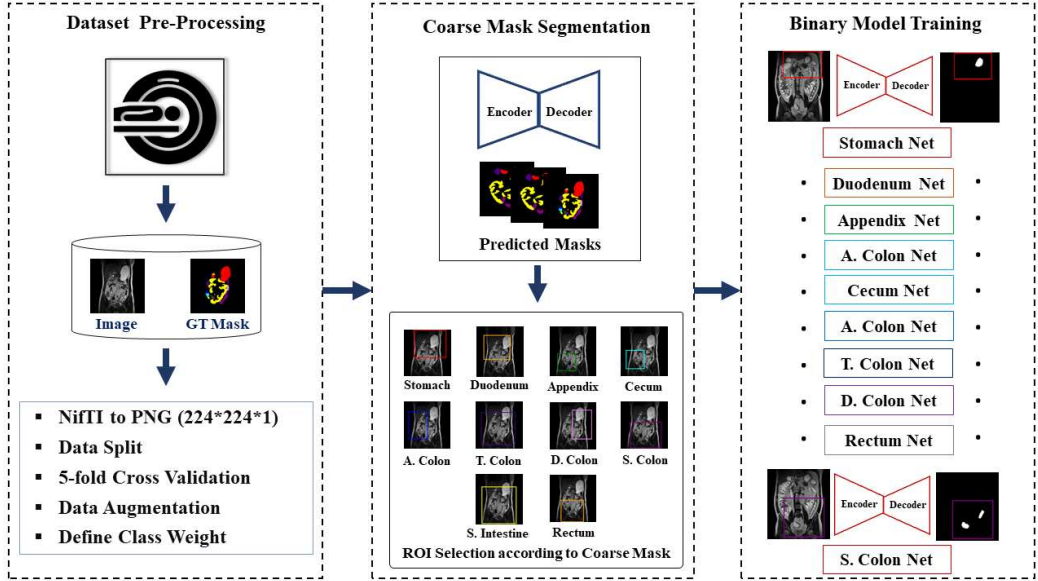

Fig. 2: Summary of the proposed framework.

### 3.1 Dataset Description

This study employed a publicly available MRE dataset designed to support intestinal segmentation research in patients with IBD [29]. The dataset consists of 3,195 coronal T2-weighted HASTE slices obtained from 114 patients diagnosed with Crohn's disease. Imaging was performed using 3.0-T Siemens Prisma and Vida scanners, following standardized bowel preparation protocols that included oral administration of mannitol and the use of anti-peristaltic agents to suppress motion artifacts. Each image in the dataset is accompanied by high-resolution, pixel-level annotations delineating ten anatomically distinct gastrointestinal structures: stomach, duodenum, small intestine, appendix, cecum, ascending colon, transverse colon, descending colon, sigmoid colon, and rectum. These annotations were generated using a deep learning-assisted labeling pipeline and subsequently refined through expert manual correction by radiologists, ensuring clinical fidelity and segmentation reliability. This dataset was chosen for its comprehensive organ-level coverage, realistic clinical acquisition conditions, and the notable rarity of publicly accessible, multi-class annotated MRE resources. It presents a meaningful benchmark for developing and validating segmentation algorithms, particularly in the context of class imbalance, anatomical overlap, and morphological heterogeneity. Nonetheless, the dataset has inherent limitations. Notably, the jejunum and ileum are not separately annotated due to their indistinct appearance in coronal MRE views. Additionally, structures such as the appendix and cecum exhibit greater segmentation difficulty across methods, often resulting in lower Dice scores, reflecting both their small size and anatomical ambiguity. Despite these challenges, this dataset remains a critical foundation for advancing the state of the art in gastrointestinal image analysis.

### 3.2 Dataset Preprocessing

The original MRE datasets, comprising 3D volumes and segmentation labels in NIfTI format, were first converted into 2D grayscale PNG slices to facilitate compatibility with standard convolutional models. For the first-stage coarse segmentation task, multiclass ground truth masks were generated using indexed palette coloring to represent all ten gastrointestinal structures within a single image. To prepare data for the second-stage organ-specific models, class-wise preprocessing was employed. This included extracting 2D ROI around each organ based on the predictions from the first stage, thereby enabling focused refinement in anatomically relevant areas. Each ROI slice was resized to a uniform resolution and paired with its respective binary mask. Class imbalance was addressed through class-wise weighting in the loss function, computed based on pixel frequency across the training set. Additionally, data augmentation strategies were applied separately for each organ class to enhance inter-class variability, particularly for underrepresented structures. While this preprocessing pipeline introduces simplification through 2D conversion and resizing, it ensures standardized input, balanced learning across classes, and refined localization for both global and organ-specific segmentation tasks.

### 3.3 Data Augmentation

To enhance the diversity of the training data and improve model generalization, we applied a set of data augmentation techniques during training. These included both geometric and intensity-based transformations, implemented on-the-fly to maintain alignment between input images and corresponding ground truth masks. Specifically, the geometric augmentations included random rotations in the range of –20° to +20° and affine shear transformations within –2° to +2°, simulating anatomical variability and local distortions common in abdominal imaging. In addition, random brightness and contrast adjustments were applied within a range of ±20% to account for signal intensity variability across MRE scans. These augmentations were particularly important for improving model robustness in underrepresented classes and complex structures with variable appearances.

### 3.4 Multi-Class Segmentation Model Architecture

To perform coarse segmentation of GI organs from MRE slices, we employed a modified UNet++ architecture with a pretrained DenseNet201 backbone, as illustrated in Fig. 3 [30], [31]. This architecture integrates the efficient feature reuse of DenseNet201 with the multi-scale skip connectivity and deep supervision capabilities of UNet++, offering strong representational capacity for complex abdominal segmentation tasks.

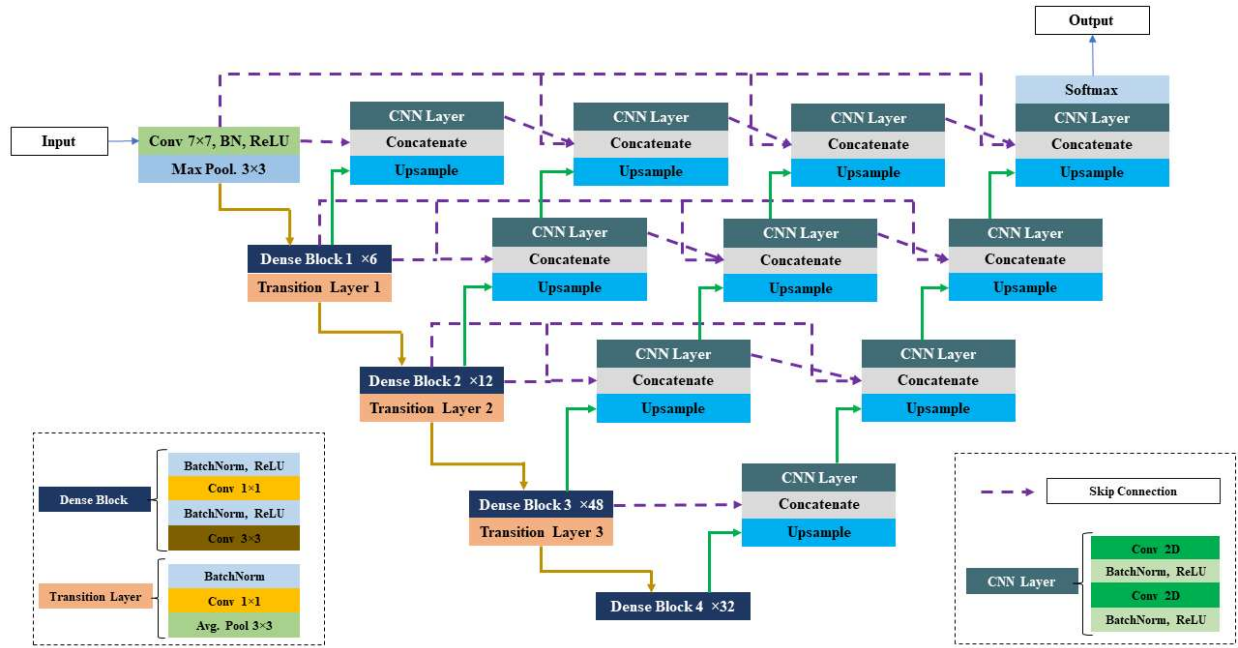

Fig. 3: Architecture of initial stage multi-class segmentation model, DenseNet201-UNet++.

The encoder leverages the DenseNet201 backbone, which comprises densely connected convolutional layers grouped into four dense blocks, interleaved with transition layers. Each dense block promotes gradient flow and feature reuse by connecting each layer to every subsequent layer within the block. Transition layers use batch normalization, 1×1 convolutions, and 2×2 average pooling to progressively reduce spatial resolution while maintaining rich semantic information. The encoder begins with a 7×7 convolution followed by batch normalization, ReLU activation, and a 3×3 max pooling operation.

The decoder follows the UNet++ nested skip connectivity structure. Each decoder node receives concatenated features from the corresponding encoder output and from all prior decoder stages at the same or deeper levels. This nested design facilitates multi-scale feature fusion, reduces the semantic gap between encoder and decoder paths, and enhances spatial precision. All convolutional layers in the decoder use standard 3×3 convolutions, followed by batch normalization and ReLU activations. At each decoder level, feature maps are upsampled using bilinear interpolation and concatenated with encoder outputs and preceding decoder features. This design encourages dense aggregation of semantic and spatial cues across multiple resolutions, improving accuracy in segmenting both large and small anatomical regions. The final decoder output is passed through a 1×1 convolutional layer followed by a softmax activation, producing a multi-class probability map where each channel corresponds to one of the ten distinct annotated GI organs.

### 3.4 Region-of-Interest (ROIs) Selection

To enable anatomically consistent and accurate refinement in the second stage of our segmentation framework, we implemented a volumetric ROI extraction strategy based on the coarse predictions generated by the initial multi-class DenseNet201–U-Net++ model. As illustrated in Fig. 4, we reconstructed both the original MRE slices and their corresponding predicted segmentation masks into 3D volumes by stacking 2D slices for each patient. From each 3D prediction volume, we created class-wise binary masks by isolating the relevant anatomical labels. We then computed tight 3D bounding boxes around non-zero voxels within each binary mask and expanded them using a fixed 40-pixel padding in all directions. These bounding boxes were then used to extract the corresponding ROIs from the original MRE volumes. Finally, each ROI volume was resliced into 2D patches, intensity-normalized, and resized to match the input resolution of the second-stage binary segmentation models.

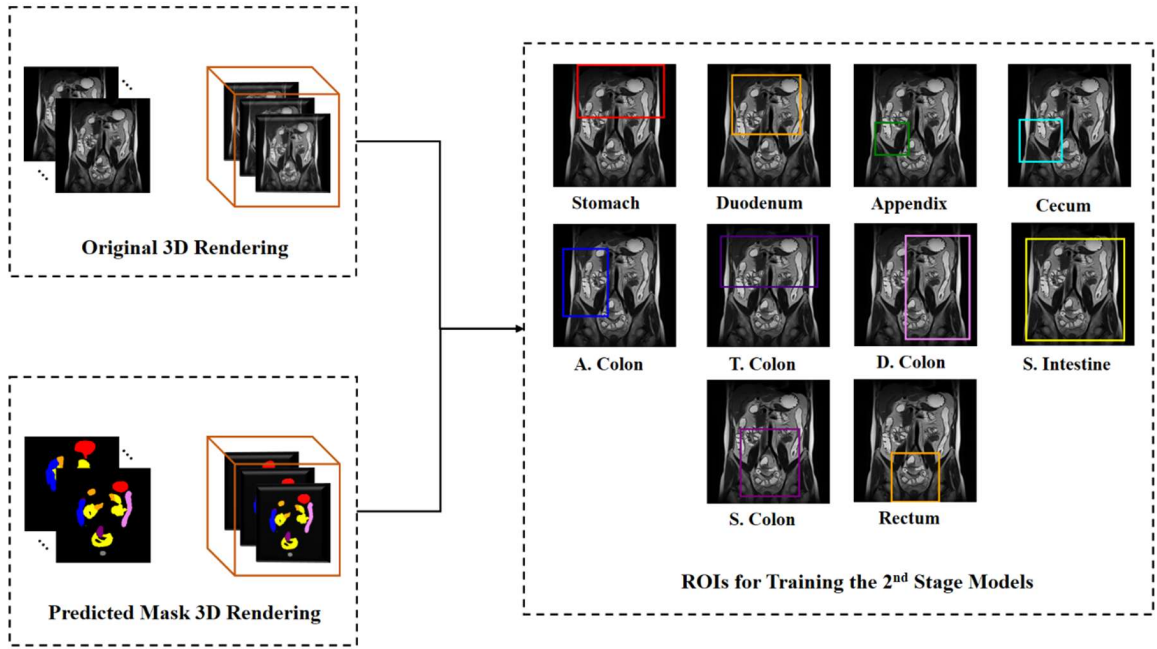

Fig. 4: Volumetric ROIs were extracted by stacking 2D predictions into 3D segmentation volumes. Tight bounding boxes with 40-pixel padding were computed for each organ to localize ROIs in the original MRE. These localized patches were used as input for second-stage organ-specific binary segmentation. Example shown for subject 72, slice 15.

This volumetric approach was designed to address the limitations of conventional 2D slice-wise segmentation. In 2D workflows, we observed that predictions were often inconsistent across adjacent slices, organs might appear clearly segmented in one slice but be missing or fragmented in the next. These discontinuities pose a risk of losing clinically relevant slices during ROI selection, particularly for small or low-contrast organs such as the appendix, rectum, or sigmoid colon. By reconstructing predictions into a 3D volume, we preserved spatial continuity across slices and treated each organ as a coherent anatomical structure. This ensured that even when segmentations by initial multi-class model were sparse or fragmented, the full spatial extent of the organ was captured and reliably passed on to the second-stage binary segmentation model for refinement.

The 40-pixel padding applied around each bounding box served two key purposes. First, the padding preserved essential anatomical context near the organ boundaries, which is often underrepresented in coarse segmentation outputs due to limited spatial resolution or class imbalance in the first-stage model. Second, it accommodated minor localization errors and partial volume effects, ensuring that the extracted ROIs included the complete organ. This padding size was determined empirically by evaluating several values (ranging from 20 to 60 pixels) on a development subset. A 40-pixel margin consistently offered the best balance between including sufficient context and avoiding unnecessary background. Manual inspection confirmed that this setting reliably enclosed each target structure across a range of anatomical variations. By converting each padded 3D ROIs defined images into a set of normalized 2D slices, we ensured compatibility with the second-stage model architecture while maintaining spatial focus on the region of interest.

### 3.6 Class-Wise Binary Segmentation Model Architecture

To enhance segmentation accuracy for GI organs that are small, low-contrast, or morphologically complex, we introduce class-wise binary segmentation strategy as the second stage of our proposed framework. In this stage, individual models are trained separately for each organ using focused ROI patches derived from the coarse segmentation outputs of the first-stage multi-class model. The architecture of each binary segmentation model is illustrated in Fig. 5. It adopts a symmetric encoder–decoder structure composed of a truncated DenseNet121 encoder and a SelfONN-based decoder, connected via skip pathways.

Each model follows a symmetric encoder–decoder design, where the encoder is derived from a truncated version of DenseNet121 and the decoder is composed of SelfONN layers. The encoder consists of four dense blocks with progressively increasing depth, separated by transition layers that downsample the feature maps. Within each dense block, 3×3 convolutional units are interleaved with batch normalization and ReLU activation functions. The use of dense connectivity enables feature reuse and stabilizes gradient propagation across the network. The input image first passes through a 7×7 convolution layer followed by batch normalization and a 3×3 max pooling layer, which together capture early spatial representations before the downsampling stages begin.

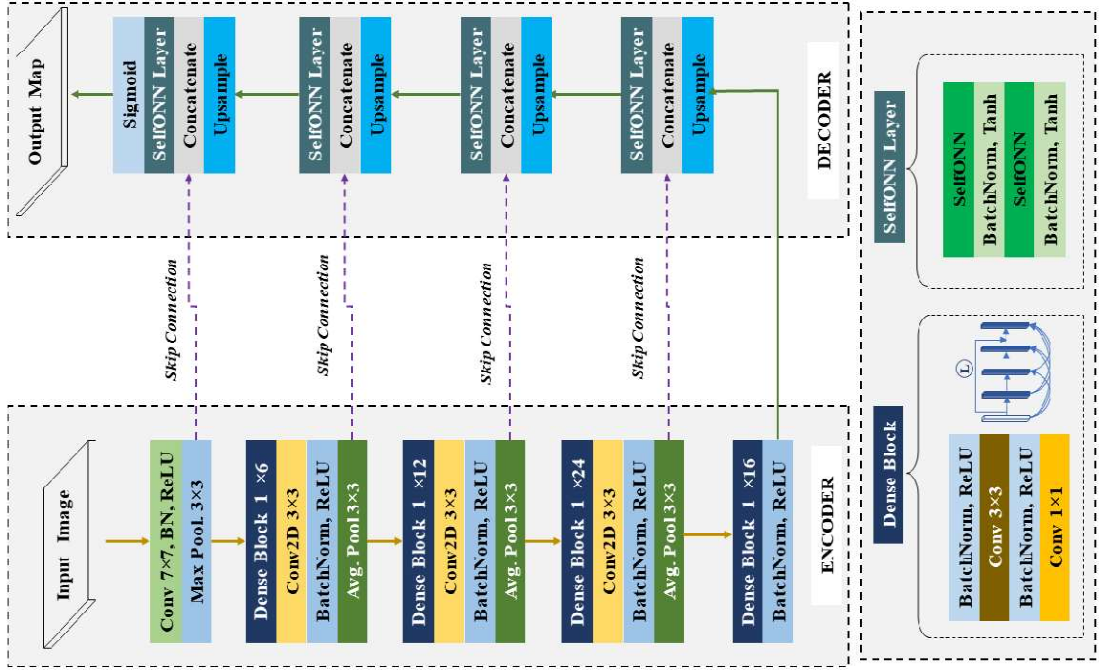

Fig. 5: Architecture of the proposed novel class-wise binary segmentation model.

The decoder mirrors the hierarchical structure of the encoder and is responsible for reconstructing high-resolution segmentation masks. Upsampling is performed at each decoder level, followed by a SelfONN layer, which replaces traditional convolutional operations with nonlinear, polynomial-based transformation units. Each SelfONN neuron is modeled as a learnable Taylor series expansion (Eq. 5), enabling it to dynamically approximate complex mappings suitable for the target organ's texture and morphology. At every level, skip connections link the encoder and decoder pathways, allowing fine-grained spatial information from the encoder to be fused with semantic features in the decoder. This skip-based integration supports both boundary refinement and contextual consistency. The final decoder output is passed through one additional SelfONN layer, followed by a sigmoid activation function to produce a binary probability map for the target class. This output is then resized and mapped back into the full-resolution image domain for spatial alignment with the original MRE slice.

This class-wise design leverages the representational strength of DenseNet121 for hierarchical feature encoding and the nonlinear flexibility of SelfONNs for spatial refinement. As a result, the model effectively improves segmentation performance in targeted classes that are morphologically complex or poorly defined in MRE scans, such as the appendix, sigmoid colon, and cecum.

## 3.7 SelfONN

Self-Organized Operational Neural Networks (SelfONNs) build upon the foundation of Operational Neural Networks (ONNs) by introducing generative neurons that can autonomously learn and optimize their nodal operators during training [32]. This dynamic adaptation enhances both the flexibility and computational efficiency of the network, as each neuron is capable of generating and refining a custom combination of nonlinear operators tailored to the task. In our previous study, incorporating SelfONNs in the decoder stage demonstrated improved segmentation performance, particularly in capturing fine anatomical boundaries [33].

In conventional Convolutional Neural Networks (CNNs), the activation of the k-th neuron in layer l is computed through a linear convolution followed by a fixed nonlinearity, described in Eq. 1.

$$x_k^l = b_k^l + \sum_{i=1}^{N_{l-1}} \text{conv } 2D(w_{ki}^l, y_i^{l-1})  \qquad 1$$

At the pixel level, this operation can be expanded as:

$$x_k^l(m,n)\Big|_{(0,0)}^{(M-1,N-1)} = \sum_{r=0}^{2}\sum_{t=0}^{2}\left(w_{ki}^l(r,t)y_i^{l-1}(m+r,n+t)\right) + \cdots \qquad 2$$

Here, $w_{ki}^l$ denotes the convolution kernel, and $y_i^{l-1}$ is the input feature map from the previous layer. ONNs extend this model by replacing the fixed activation function with learnable nodal operators Ψ and incorporating pooling operators P, giving each neuron the capacity to apply a wider range of nonlinear transformations, described in Eq. 3.

$$x_k^l = b_k^l + \sum_{i=1}^{N_{l-1}} \text{oper } 2D(w_{ki}^l, y_i^{l-1}) \qquad 3$$

This is further expressed in spatial form as:

$$x_k^l(m,n)\big|_{(0,0)}^{(M-1,N-1)} = b_k^l + \sum_{i=1}^{N_{l-1}} \left( P_k^l \begin{bmatrix} \Psi_{ki}^l(w_{ki}^l(0,0), y_i^{l-1}(m,n)), \dots, \\ \Psi_{ki}^l(w_{ki}^l(r,t), y_i^{l-1}(m+r, n+t), \dots), \dots \end{bmatrix} \right) \qquad 4$$

While ONNs improve expressiveness by allowing nonlinearity customization, they still depend on a predefined library of nodal functions. SelfONNs overcome this constraint by allowing each nodal operator to be learned from scratch using a Taylor series-based expansion. Specifically, the operator is formulated as a polynomial of order Q as described in Eq. 5.

$$\Psi(w, y) = w_0 + w_1 y + w_2 y^2 + \cdots + w_Q y^Q \qquad 5$$

During forward propagation, each kernel element's nodal operator in Self-ONNs is approximated by this composite nodal operator:

$$\Psi(w, y_{(m+r,n+t)}) = w_{r,t,0} + w_{r,t,1} y_{(m+r,n+t)} + w_{r,t,2} y_{(m+r,n+t)}^2 + \cdots + w_{r,t,Q} y_{(m+r,n+t)}^Q \qquad 6$$

This formulation enables each neuron to self-organize its nodal behavior, continuously optimizing it during training to best fit the input distribution. As a result, SelfONNs demonstrate improved learning capacity and computational efficiency, often outperforming both ONNs and CNNs—even when using more compact architectures.

### 3.8 Experiment Setup

The proposed dual-stage segmentation pipeline was evaluated through systematic experiments on a curated dataset of abdominal MRE slices. This section outlines the implementation environment, data preparation strategy, and training configurations for both the initial multiclass segmentation model and the second-stage class-wise binary models.

### 3.8.1 Initial Multiclass Segmentation Model Training Setup

The initial multi-class segmentation model, DenseNet201-UNet++, was trained to perform pixel-wise multiclass segmentation of ten distinct GI organs. Input slices were resized to 224×224 and normalized to zero mean and unit variance. To ensure robust performance evaluation and generalizability, we adopted a 5-fold cross-validation approach. In each fold, the dataset was split into 70% training, 10% validation, and 20% testing, maintaining patient-wise stratification to preserve proportional representation and eliminate data leakage across sets. The details model training configuration is summarized in Table 1.

Table 1. Summary of multiclass coarse segmentation model training configuration.

| Training Components | Specifications |
|---|---|
| Input Size | 224 × 224 × 1 |
| Loss Function | CrossEntropy |
| Data Augmentation | Flip, Rotation, Brightness/Contrast adjustment |
| Optimizer | Adam |
| Initial Learning Rate (LR) | 1e-4 |
| LR Scheduler | ReduceLROnPlateau (factor = 0.5, patience = 5) |
| Batch Size | 16 |
| Epochs | 100 (early stopping, patience = 20) |
| Dataset Split | 70% train / 10% val / 20% test (patient-stratified) |
| Cross Validation | 5-Fold |
| Framework & Enviroment | PyTorch 2.0, Python 3.11, CUDA 12.1 |
| Hardware | NVIDIA RTX 4090 (24 GB) |

To address severe class imbalance especially for small structures like the appendix a class-weighted CrossEntropy loss was used. Training was conducted using the Adam optimizer with an initial learning rate of 1e-4, adjusted using a ReduceLROnPlateau scheduler (factor = 0.5, patience a score of 5). Early stopping (patience a score of 20) was applied based on validation loss. Training proceeded for up to 100 epochs, with early stopping (patience a score of 20) based on validation loss. A batch size of 16 was selected to balance memory constraints of the high-resolution input and the large model size while maintaining stable gradient update. Augmentation included random rotation, flipping, elastic deformation, and contrast adjustment. All training was performed in PyTorch 2.0 with CUDA 12.1, using Python 3.11, on an NVIDIA RTX 4090 GPU 24 GB.

### 3.8.2 Class-Wise Binary Segmentation Models Training Setup

The second-stage models were trained independently for each organ using the ROIs from the initial segmentation model. All organ-specific models followed a DenseNet121–SelfONN U-Net-style architecture, where SelfONN layers also used a Q-order of 3.

The training loss was defined as the mean of Dice loss and Jaccard loss. A 5-fold cross-validation strategy was employed using the same patient-wise stratification protocol as in the first stage to ensure robust and generalizable performance. Training was conducted using the Adam optimizer with an initial learning rate of 1e-4, which was adaptively reduced via a ReduceLROnPlateau scheduler (factor = 0.5, patience a score of 20). Each fold was trained for up to 100 epochs with early stopping based on validation loss. A batch size of 16 was used. Data augmentation included random flips, small-angle rotations, and brightness–contrast variations. All models were implemented in PyTorch 2.0, running on Python 3.11 and CUDA 12.1, using an NVIDIA RTX 4090 GPU. Key settings are summarized in Table 2.

Table 2. Summary of proposed binary class-wise model training configuration.

| Training components | Specification |
|---|---|
| Input Size | Padded ROIs from Stage 1 Coarse Model |
| Loss Function | Mean of Dice Loss and Jaccard Loss |
| SelfONN Q-order | 3 |
| Optimizer | Adam |
| Initial Learning Rate (LR) | 1e-4 |
| LR Scheduler | ReduceLROnPlateau (factor = 0.5, patience = 15) |
| Batch Size | 16 |
| Epoch | 100 (early stopping, patience = 20) |
| Dataset Split | 5 folds, 70% train, 10% val , and 20% test (patient-wise split) |
| Augmentation | Flip, Rotation, Brightness/Contrast adjustment |
| Framework & Env. | PyTorch 2.0, Python 3.11, CUDA 12.1 |
| Hardware | NVIDIA RTX 4090 24 GB |

The choice of hyperparameters was guided by both empirical performance and practical constraints. The mean of Dice and Jaccard losses was chosen to enforce a balance between pixel-level accuracy and overall region-level overlap, especially important for irregular or small organs. The Adam optimizer is well-suited for segmentation tasks due to its adaptive learning capabilities, and an initial learning rate of 1e-4 offers a reliable trade-off between convergence speed and stability. The ReduceLROnPlateau scheduler helps adapt learning when progress stalls, making training more resilient. A batch size of 16 was selected to ensure efficient GPU utilization without exceeding memory limits, given the high resolution and model size. Early stopping (patience a score of 20) helped prevent overfitting, particularly in folds with limited validation samples. Lastly, data augmentation techniques such as flipping, rotation, and intensity variation were introduced to simulate anatomical variability and improve generalization across unseen cases.

### 3.9 Loss Function

To optimize the performance of the proposed dual-stage segmentation pipeline, we employed distinct loss functions tailored to the nature of each task: CrossEntropy loss for the first-stage multiclass segmentation and a combined Dice–Jaccard loss for the second-stage organ-wise binary segmentation. These choices were motivated by the need to handle class imbalance, promote boundary accuracy, and ensure region-level agreement between prediction and ground truth.

In the first-stage multiclass segmentation model, we used CrossEntropy loss, a standard choice for pixel-wise classification problems involving multiple classes [34]. It measures the divergence between the predicted class probabilities and the true one-hot encoded labels. For an image with N pixels and C classes, the CrossEntropy loss is expressed in Eq. 6.

$$\mathcal{L}_{CE} = -\frac{1}{N}\sum_{i=1}^{N}\sum_{c=1}^{C} y_{i,c} \log(p_{i,c}) \qquad 6$$

Where $y_{i,c} \in [0,1]$ is the ground truth label for class c at pixel i, and $p_{i,c} \in [0,1]$ is the predicted softmax probability for class ccc at the same pixel.

In this study, CrossEntropy was chosen for its ability to directly supervise dense classification across multiple organ classes, including structures with overlapping or indistinct borders. To address the class imbalance particularly for the appendix, which is small and infrequently visible are assigned as a class weight seven times higher than the other classes. This weighting penalized false negatives more heavily and encouraged the model to learn finer features for under-represented organs.

For the second-stage binary segmentation models, each organ-specific model was trained using a composite loss function combining Dice loss and Jaccard loss (also known as IoU loss). These region-based metrics are well-suited for binary tasks involving foreground–background discrimination and are particularly effective for imbalanced foreground classes.

The Dice loss, derived from the Dice Similarity Coefficient, emphasizes overlap between predicted and ground truth masks [35]. It is presented mathematically in Eq. 7.

$$\mathcal{L}_{Dice} = 1 - \frac{2\sum_{i=1}^{N} p_i, y_i + \epsilon}{\sum_{i=1}^{N} p_i + \sum_{i=1}^{N} y_i + \epsilon} \qquad 7$$

The Jaccard loss (IoU loss) is more stringent and penalizes over or under-segmentation by considering the ratio of intersection to union [35], mathematically expressed in Eq. 8.

$$\mathcal{L}_{Jaccard} = 1 - \frac{2\sum_{i=1}^{N} p_i, y_i + \epsilon}{\sum_{i=1}^{N}(p_i + y_i - p_i, y_i) + \epsilon} \qquad 8$$

Here, $p_i \in \{0,1\}$ is the predicted probability and $y_i \in \{0,1\}$ is the ground truth at pixel i, and $\epsilon$ is a small constant to prevent division by zero.

The final objective used for second-stage training is the average of both Dice and Jaccard loss as expressed in Eq. 9.

$$\mathcal{L}_{final} = \frac{1}{2}(\mathcal{L}_{Dice} + \mathcal{L}_{Jaccard}) \qquad 9$$

This hybrid formulation provides a balanced optimization signal, capturing both region-level accuracy (via Jaccard) and shape sensitivity (via Dice). In this study, it was especially beneficial for refining the segmentation of small, irregular organs such as the cecum, appendix, and sigmoid colon, which suffer from fuzzy boundaries and frequent under-segmentation when trained with classification-only objectives.

By strategically selecting these loss functions, our approach ensured robust training performance in both multiclass and binary settings, while mitigating the effects of class imbalance and morphological variability.

### 3.10 Evaluation Metrics

To comprehensively assess the segmentation performance of the proposed framework, we employ three widely used metrics: Dice Similarity Coefficient (DSC), Intersection over Union (IoU), and 95th Percentile Hausdorff Distance (HD95). These metrics are chosen to evaluate both region-level agreement and boundary-level accuracy across diverse abdominal organ structures.

#### 3.10.1 Dice Similarity Coefficient (DSC)

DSC measures the overlap between the predicted segmentation P and the ground truth G [36], and is defined as,

$$\text{DSC} = \frac{2|P \cap G|}{|P| + |G|} \qquad 10$$

It ranges from 0 to 1, with higher values indicating better spatial agreement. In this study, DSC serves as the primary evaluation metric for both stages, as it balances precision and recall, and is particularly effective in assessing organ-level segmentation performance, even in cases of class imbalance.

#### 3.10.2 Intersection over Union (IoU)

Also known as the Jaccard Index [37], IoU quantifies the ratio of the intersection to the union of predicted and ground truth regions and it is defined below,

$$\text{IoU} = \frac{|P \cap G|}{|P \cup G|} \qquad 11$$

IoU is more conservative than DSC and penalizes both under- and over-segmentation. It provides complementary insight into overall region accuracy, and is especially useful for validating segmentation quality on overlapping or irregularly shaped organs.

#### 3.10.3 95$^{th}$ Percentile Hausdorff Distance (HD95)

HD95 measures the distance between the boundary points of the predicted segmentation mask P and the ground truth mask G [37], excluding the top 5% of outlier distances. This metric provides a robust evaluation of boundary alignment by focusing on the bulk of the shape contour, while disregarding extreme deviations caused by noise or small annotation errors. Mathematically, HD95 is defined as,

$$HD95(P, G) = \max \left\{ \begin{array}{l} \text{percentile}_{95} \left( \min_{g \in G} \|p - g\| \right)_{p \in P}, \\ \text{percentile}_{95} \left( \min_{p \in P} \|g - p\| \right)_{g \in G} \end{array} \right\} \quad 12$$

P and G are the sets of surface points of the predicted and ground truth masks, respectively. $\|p-g\|$ denotes the Euclidean distance between a point $p \in P$ and $g \in G$. This metric is crucial in medical segmentation tasks where boundary accuracy can directly impact clinical interpretation. In our study, HD95 is used to evaluate anatomical boundary alignment, especially in small and critical organs where pixel-level deviations may carry high clinical significance.

By combining DSC, IoU, and HD95, we provide a balanced evaluation of segmentation overlap, completeness, and boundary precision, allowing us to rigorously assess the effectiveness of both the coarse and refined segmentation stages.

4. **Results and Discussion**

This section presents the quantitative and qualitative evaluation of the proposed dual-stage segmentation framework for segmentation of ten anatomically distinct gastrointestinal organs segmentation from MRE images. The performance of both the initial multi-class segmentation model and the second-stage class-wise refinement models is assessed to demonstrate the effectiveness of the pipeline in segmenting anatomically complex and morphologically diverse structures. The evaluation is conducted using a comprehensive set of standard metrics to reflect both region overlap and boundary accuracy.

**4.1 Initial Multiclass Segmentation Model Performance**

In the first stage of our proposed two-stage segmentation framework, a series of multiclass segmentation models was trained to generate coarse masks from MRE slices. The primary objective of this stage was to identify the most effective model for accurate organ-wise localization, which would serve as the basis for ROI extraction in the subsequent organ-wise refinement phase.

To evaluate benchmark performance, several state-of-the-art encoders–decoder architectures, including ResNet50–UNet, ResNet50–UNet++, EfficientNetB4–UNet++, VGG19–UNet++, and DenseNet201–UNet++ models were implemented and trained on ROI images. All models were trained under identical conditions, using the same preprocessing pipeline, loss function, and cross-validation protocol to ensure fair comparison. Segmentation performance was assessed on a held-out test set using the DSC and HD95, evaluated across ten abdominal organ classes.

As presented in Table 3, the proposed DenseNet201–UNet++ model consistently achieved the highest segmentation accuracy and boundary precision across most organ classes. The model showed strong performance on large, high-contrast structures such as the stomach, small intestine, and rectum, while also outperforming the baselines on more morphologically variable regions like the sigmoid colon and transverse colon. Among other models, EfficientNetB4–UNet++ performed comparatively well, particularly in terms of HD95 in boundary-sensitive organs. However, its performance was less consistent across classes. ResNet50–UNet++ demonstrated moderate gains over the standard UNet configuration, while VGG19–UNet++ exhibited limited robustness, particularly in deeper abdominal segments.

Table 3. Performance of multi-class segmentation model at initial stage.

| Models | ResNet50–UNet | | ResNet50–UNet++ | | Efficientb4–UNet++ | | VGG19–UNet++ | | DenseNet201–UNet++ | |
|---|---|---|---|---|---|---|---|---|---|---|
| Metrics | DSC | HD95 | DSC | HD95 | DSC | HD95 | DSC | HD95 | DSC | HD95 |
| Stomach | 81.55 | 7.9 | 87.09 | 7.63 | 89.10 | 5.21 | 85.20 | 6.41 | 94.60 | 4.46 |
| Duodenum | 77.41 | 10.02 | 79.98 | 9.95 | 78.32 | 8.57 | 78.62 | 8.57 | 85.22 | 6.95 |
| S. Intestine | 72.84 | 10.91 | 86.17 | 6.91 | 83.87 | 8.79 | 78.10 | 11.19 | 90.17 | 7.19 |
| Appendix | -- | -- | 6.21 | -- | 5.07 | -- | -- | -- | 6.76 | -- |
| Cecum | 55.42 | 12.12 | 67.43 | 7.59 | 68.27 | 7.69 | 68.03 | 7.16 | 69.40 | 7.50 |
| Asc. Colon | 67.98 | 12.56 | 85.41 | 8.07 | 81.23 | 9.06 | 77.23 | 10.73 | 85.41 | 8.07 |
| Trans. Colon | 71.9 | 18.44 | 71.83 | 21.25 | 78.18 | 16.54 | 71.08 | 17.54 | 84.08 | 16.25 |
| Desc. Colon | 72.31 | 13.44 | 82.48 | 11.01 | 82.45 | 11.24 | 81.54 | 11.94 | 85.54 | 11.13 |
| Sig. Colon | 52.94 | 12.14 | 79.44 | 11.29 | 81.29 | 10.42 | 82.71 | 10.02 | 84.08 | 9.92 |
| Rectum | 57.00 | 8.64 | 85.15 | 4.90 | 83.84 | 4.47 | 81.83 | 4.77 | 86.83 | 3.97 |

A significant challenge emerged in the segmentation of the appendix. During the study, the model encountered a severe class imbalance problem which affects the organ. Among the 114 patients included in the dataset, the appendix was annotated in only 63 patients, and within those, it appeared in just 258 out of 1,666 slices. Moreover, the appendix occupied a very small spatial area, often comprising only a few pixels with low contrast relative to surrounding structures. Its indistinct boundaries and proximity to other intestinal loops further complicated its detection by convolutional models.

Despite applying extensive data augmentation techniques and experimenting with class reweighting strategies, the segmentation performance for the appendix remained unreliable across all evaluated models, including DenseNet201–UNet++. To address this limitation, in the following section, the best-performing model with class-specific weighting was trained to mitigate the effects of class imbalance. Then the performance of the DenseNet201–UNet++ model was compared with and without class weighting, with particular focus on its impact on appendix segmentation.

### 4.1.1 Effect of Class Weighting on Overall Performance

To address the severe class imbalance observed in the appendix class, the DenseNet201–UNet++ model was trained using a class-weighted cross-entropy loss. A significantly higher class weight was assigned to the appendix to encourage the model to better learn from its sparse and small-scale representation in the training data. The performance of the weighted model was then compared against the unweighted version across all organ classes using DSC, IoU, and HD95, as summarized in Table 3.

Table 4. Effect of class weighting on overall performance of initial stage DenseNet201–UNet++ model.

| Class | Without Class Weight | | | Class Weight | | | | | |
|---|---|---|---|---|---|---|---|---|---|
| | DSC | IoU | HD95 | DSC | Δ DSC | IoU | Δ IoU | HD95 | Δ HD95 |
| **Stomach** | 94.60 | 89.84 | 4.46 | 85.53 | -9.07 | 83.74 | -6.1 | 7.36 | +2.9 |
| **Duodenum** | 85.22 | 83.03 | 6.95 | 80.23 | -4.99 | 77.13 | -5.9 | 10.94 | +3.99 |
| **S. Intestine** | 90.17 | 82.47 | 7.19 | 75.11 | -15.06 | 68.24 | -14.23 | 11.19 | +4.37 |
| **Appendix** | 6.76 | 4.07 | -- | 85.76* | +79.05* | 82.07* | +78.05* | 2.01* | +2.01* |
| **Cecum** | 69.40 | 65.77 | 7.50 | 59.4 | -10 | 54.77 | -11 | 12.53 | +5.03 |
| **Asc. Colon** | 85.41 | 83.76 | 8.07 | 71.76 | -13.65 | 66.41 | -17.35 | 14.07 | +6.9 |
| **Trans. Colon** | 84.08 | 74.04 | 16.25 | 74.04 | -10.04 | 64.09 | -9.95 | 21.25 | +5.42 |
| **Desc. Colon** | 85.54 | 82.13 | 11.13 | 73.74 | -11.8 | 71.33 | -10.8 | 16.13 | +5.1 |
| **Sig. Colon** | 84.08 | 80.49 | 9.92 | 71.08 | -12.9 | 66.49 | -36 | 13.92 | +4.06 |
| **Rectum** | 86.83 | 77.38 | 3.97 | 73.83 | -13.02 | 69.38 | -23 | 9.97 | +5.81 |

The results revealed a substantial improvement in appendix segmentation. With class weighting, the DSC increased from 6.76% to 85.76%, and IoU improved from 4.07% to 82.07%, while HD95 shown 2.01 mm. These improvements demonstrated the effectiveness of class reweighting in recovering performance for underrepresented and small-sized structures that were otherwise neglected during optimization.

However, the inclusion of aggressive class weights introduced trade-offs. Most other organ classes exhibited a decline in performance across all metrics. For instance, DSC dropped by more than 13% in the ascending colon, 15% in the small intestine, and nearly 9% in the stomach. Similarly, HD95 values increased across all classes, indicating reduced boundary accuracy. This suggests that overcompensating for a rare class may distort the optimization landscape, leading to compromised segmentation quality in larger or more prevalent structures.

Despite these trade-offs, we selected the class-weighted DenseNet201–UNet++ model as the coarse segmentation model for initial stage, owing to its ability to generate anatomically complete organ masks, including the appendix. Subsequent refinements and segmentation accuracy improvements were addressed in second stage through organ-specific models, trained on localized ROI patches.

### 4.1.2 Performance of Class-wise Binary Models

To improve segmentation accuracy particularly for small, low-contrast, morphologically complex abdominal organs, a novel refinement strategy using dedicated novel DenseNet121–SelfONN U-Net models was developed which was trained on ROI patches. These ROIs were derived from the coarse segmentation masks generated by the first-stage DenseNet201–UNet++ model.

As shown in Table 5, the second stage proposed models demonstrated significant improvements across almost all organ classes when compared to the initial stage coarse segmentation results. The DSC improved substantially for instance, by 23.62% in the cecum, 18.57% in the sigmoid colon, and 17.99% in the rectum. Similarly, IoU scores increased accordingly, and the HD95 decreased consistently, indicating enhanced boundary precision. Notably, the transverse colon and ascending colon exhibited large HD95 reductions (–9.06 mm and –5.36 mm, respectively), suggesting that the organ-wise refinement step effectively corrected coarse boundary predictions.

For the appendix, where class weighting had already yielded a large gain in initial stage, the second stage model still produced further improvements in both overlap and marginally in boundary alignment. These results validated the benefit of using localized models with task-adaptive nonlinear decoding layers to fine-tune segmentation, particularly for anatomically ambiguous or underrepresented classes.

Table 5. Segmentation performance comparison between initial multi-class (DenseNet201–UNet++ with class weighting) and second stage (DenseNet121–SelfONN–UNet) models for individual GI organs.

| Class | Initial Stage Model | | | Second Stage Model | | | | | |
| --- | --- | --- | --- | --- | --- | --- | --- | --- | --- |
| | DSC | IoU | HD95 | DSC | Δ DSC | IoU | Δ IoU | HD95 | Δ HD95 |
| **Stomach** | 85.53 | 83.74 | 7.36 | 96.87 | 11.34 | 93.8 | 10.06 | 4.26 | -3.1 |
| **Duodenum** | 80.23 | 77.13 | 10.94 | 86.22 | 5.99 | 83.03 | 5.9 | 5.85 | -5.09 |
| **S. Intestine** | 75.11 | 68.24 | 11.19 | 91.17 | 16.06 | 83.47 | 15.23 | 7.19 | -4 |
| **Appendix** | 85.76* | 82.07* | 2.01* | 89.75 | 3.99 | 87.66 | 5.59 | 2.32 | +0.31 |
| **Cecum** | 59.4 | 54.77 | 12.53 | 83.02 | 23.62 | 79.78 | 25.01 | 7.21 | -5.32 |
| **Asc. Colon** | 71.76 | 66.41 | 14.07 | 87.41 | 15.65 | 83.76 | 17.35 | 8.71 | -5.36 |
| **Trans. Colon** | 74.04 | 64.09 | 21.25 | 86.54 | 12.5 | 77.04 | 12.95 | 12.19 | -9.06 |
| **Desc. Colon** | 73.74 | 71.33 | 16.13 | 87.54 | 13.8 | 85.13 | 13.8 | 10.33 | -5.8 |
| **Sig. Colon** | 71.08 | 66.49 | 13.92 | 89.65 | 18.57 | 86.57 | 20.08 | 7.98 | -5.94 |
| **Rectum** | 73.83 | 69.38 | 9.97 | 91.82 | 17.99 | 87.38 | 18.0 | 3.35 | -6.62 |

A qualitative comparison in Fig. 5 illustrates the visual impact of the two-stage approach. The figure presents the ground truth (GT) mask, initial stage prediction, extracted ROI, and the refined second stage prediction for each organ. It is evident that the Stage 2 model generated more complete and precise masks across most classes. In cases like the small intestine, sigmoid colon, and rectum, the Stage 1 predictions suffered from fragmentation and under-segmentation, whereas the Stage 2 predictions restored spatial continuity and anatomical coherence. For the appendix, Stage 1 predictions were often partial or over-segmented, while the Stage 2 model was able to recover the correct structure within the localized ROI.

These findings confirm that the proposed novel DenseNet121–SelfONN UNet architecture, combined with targeted organ-wise training, serves as an effective refinement module. The integration of stage-wise processing allowed the framework to balance global localization with local precision, resulting in robust multi-organ abdominal segmentation from MRE images.

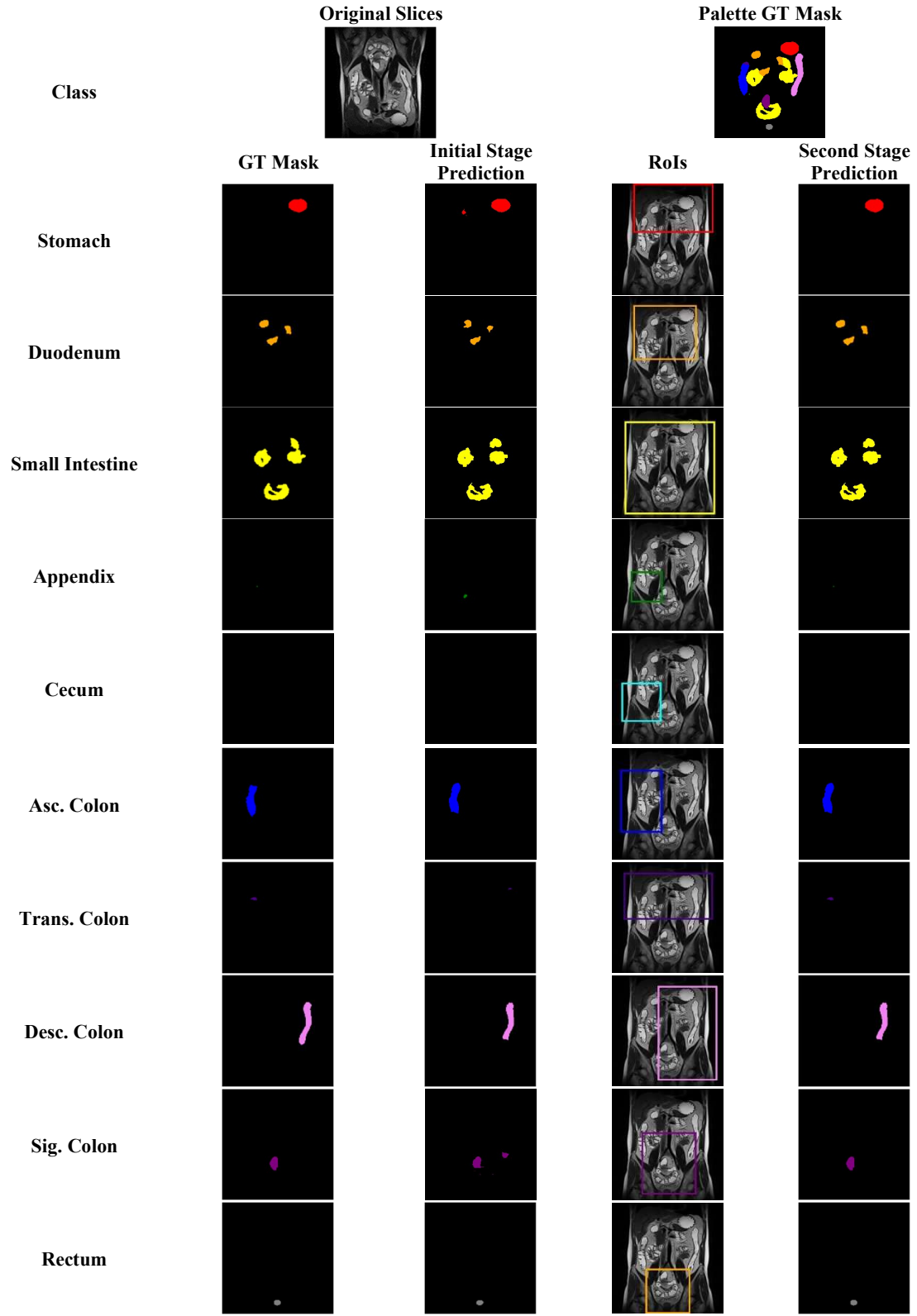

Fig. 6. Qualitative comparison of segmentation performance across ten distinct GI organs using the proposed two-stage framework. Example shown for subject 72, slice 15.

## 1.1 Benchmarking Proposed Model Against Existing Architectures

To further validate the effectiveness of the proposed DenseNet121–SelfONN U-Net model used in Stage 2, we conducted a comparative evaluation against several well-established state-of-the-art segmentation architectures. These included ResNet50–UNet, VGG19–UNet, InceptionV3–UNet, and EfficientNetB4–UNet, each representing diverse architectural backbones commonly employed in medical image segmentation. All baseline models were trained using the same preprocessed ROI patches and under identical conditions to ensure a fair and direct comparison. As summarized in Table 6, the proposed model outperformed all baseline methods across nearly every organ class in terms of both DSC and HD95. The most notable gains were observed in organs that were either morphologically complex or prone to segmentation dropouts, such as the appendix, sigmoid colon, transverse colon, and cecum. For example, the proposed model achieved a DSC of 89.75% for the appendix and 91.82% for the rectum, both surpassing all other models, and achieved the lowest HD95 values, indicating superior boundary precision.

Table 6. Comparison of existing GI structures segmentation methods with the proposed approach in terms of dataset, modality, target regions, methods, and evaluation metrics.

| Models | ResNet50–UNet | | VGG19–UNet | | InceptionV3–UNet | | EfficientB4–UNet | | Proposed Model | |
|---|---|---|---|---|---|---|---|---|---|---|
| Metrics | DSC | HD95 | DSC | HD95 | DSC | HD95 | DSC | HD95 | DSC | HD95 |
| Stomach | 93.18 | 5.23 | 91.64 | 5.78 | 94.01 | 4.66 | 95.19 | 4.44 | 96.87* | 4.26* |
| Duodenum | 82.43 | 7.49 | 81.67 | 8.03 | 84.76 | 6.86 | 85.61 | 6.28 | 86.22* | 5.85* |
| S. Intestine | 86.08 | 8.51 | 84.62 | 9.14 | 88.11 | 7.72 | 89.42 | 7.36 | 91.17* | 7.19* |
| Appendix | 84.91 | 3.84 | 82.63 | 4.49 | 87.03 | 2.93 | 88.28 | 2.68 | 89.75* | 2.32* |
| Cecum | 79.38 | 8.62 | 77.91 | 9.42 | 80.83 | 7.96 | 81.82 | 7.64 | 83.02* | 7.21* |
| Asc. Colon | 83.52 | 9.81 | 81.72 | 10.41 | 85.63 | 8.82 | 86.44 | 8.39* | 87.41* | 8.71 |
| Trans. Colon | 81.75 | 13.11 | 80.62 | 14.05 | 83.47 | 12.43 | 84.33 | 12.28 | 86.54* | 12.19* |
| Desc. Colon | 83.57 | 11.22 | 82.37 | 11.91 | 85.98 | 10.54 | 86.92 | 10.42 | 87.54* | 10.33* |
| Sig. Colon | 85.92 | 9.14 | 84.38 | 9.81 | 87.15 | 8.23 | 88.52 | 7.99 | 89.65* | 7.98* |
| Rectum | 88.71 | 4.15 | 86.14 | 4.88 | 89.66 | 3.84 | 90.87 | 3.53 | 91.82* | 3.35* |

Among the baselines, EfficientNetB4–UNet performed comparatively well in larger organs like the stomach and duodenum, reflecting its efficient representation capacity. However, it was less consistent across the remaining classes, particularly in small or anatomically ambiguous structures. InceptionV3–UNet and ResNet50–UNet showed competitive performance in mid-sized organs, though they generally trailed the proposed model in both region overlap and boundary alignment. VGG19–UNet consistently underperformed across most metrics, likely due to its limited depth and lack of dense feature reuse.

These results reinforce the strength of the proposed DenseNet121–SelfONN U-Net architecture, particularly when applied in a class-specific, ROI-focused refinement stage. Its ability to capture organ-specific nonlinearities and to generalize well across diverse anatomical structures confirms its robustness as a refinement backbone within the two-stage segmentation pipeline.

### 4.3 Comparison with Existing Literatures

To contextualize the performance and design of our proposed segmentation framework, we compare it with existing deep learning methods targeting gastrointestinal and abdominal organ segmentation across different imaging modalities. Table 7 summarizes relevant studies in terms of dataset, imaging sequence, segmented classes, methodological design, and performance metrics.

Table 7. Comparison of existing GI structures segmentation methods with the proposed approach in terms of dataset, modality, target regions, methods, and evaluation metrics.

| Ref. | Year | Dataset | Imaging Modalities | Target Regions | Methods | Key Metrics |
|---|---|---|---|---|---|---|
| [24] | 2022 | Healthy & IBD Patients | 3D cine-MRI | 1; Small Intestine | Multi-task CNN + stochastic tracker | DSC: 0.88 ± 0.03 (healthy), 0.79 ± 0.09 (IBD) |
| [38] | 2024 | 46 CD patients (202 segments) | Dual-Energy CT Enterography | 2; Intestinal segments (active/inactive CD) | Three ML models (logistic regression) based on DECTE & CT features | AUC: up to 0.87; SEN: 0.848; SPE: 0.786 |
| [26] | 2025 | 154 MRI pairs from 4 studies | T1-FS MRI, T2w MRI | 4; Colon (ascending, transverse, descending, rectum) | DEEDS registration + ICR (mesh & probabilistic modeling) | Feces coverage (R): 93.0 ± 5.2%; VOL: 11.7% |
| [22] | 2021 | 50 training + 10 testing CT volumes | CT | 1; Sigmoid colon | Iterative 2.5-D CNN with 3D+2D convolutions | DSC: 0.82 (no prior), 0.88 (with prior); HD95: 3.50 ± 2.98 mm; MSD: 1.79 mm |
| [23] | 2019 | 23 pediatric CD patients | Contrast-enhanced MRI (T1w VIBE) | 2; Lumen, Wall | CPR + 3D Residual U-Net with distance prior | DSC: Lumen 75 ± 18%, Wall 81 ± 8%, |
| [27] | 2022 | 80 abdominal MRI scans (71 patients) | T2w MRI (MR-SIM, MR-Linac) | 2; Small bowel, Large bowel, Combined bowels, Pancreas, Duodenum, Stomach | DL (3D-ResUNet) + ACR using improved level set ACM | DSC: ↑0.34→0.59 (bowels); MDA: ↓7.02→5.23 mm; APL: ↓ ~20 mm |
| Proposed Approach | | 3,195 coronal T2-W HASTE slices from 114 patients with IBD | MRE | 10; stomach, duodenum, small intestine, appendix, cecum, ascending colon, transverse colon, descending colon, sigmoid colon, and rectum | Novel Dual-Stage Deep Learning Framework | mDSC: 88.99 % mIoU: 84.76 % mHD95: 6.94 mm |

Existing deep learning methods for gastrointestinal segmentation often focus on limited anatomical regions, such as the small intestine, sigmoid colon, or select colon segments, and are typically designed for CT or conventional MRI rather than MR enterography. While approaches using multi-task CNNs, curved planar reformatting, mesh deformation, or active contour refinement have shown effectiveness in localized tasks, they commonly suffer from limited organ coverage, reliance on handcrafted post-processing, or lack of organ-specific optimization. Few studies have attempted comprehensive segmentation across diverse GI structures, and segmentation of morphologically variable or underrepresented regions such as the appendix, cecum, or sigmoid colon remains largely underexplored. Moreover, many methods depend on manual initialization or external priors, restricting automation and clinical scalability. In contrast, our proposed dual stage deep learning framework provides a fully automated, end-to-end solution tailored for MRE, enabling simultaneous segmentation of ten gastrointestinal organs. By adopting a class-aware, refinement-driven strategy, the method enhances both anatomical detail and boundary precision, while improving robustness in the presence of class imbalance and morphological variability that were commonly unaddressed in previous work.

While our approach demonstrates superior segmentation performance across ten GI organs, its advantages extend beyond numerical metrics. Unlike many existing methods that target limited regions or rely on handcrafted rules and external priors, our dual-stage framework enables comprehensive segmentation across anatomically diverse GI structures, including morphologically complex or underrepresented organs like the appendix and cecum. By decoupling global localization from organ-specific refinement, the pipeline allows each structure to be modeled with task-specific focus, improving precision and robustness, particularly in the presence of class imbalance or ambiguous anatomy. The use of SelfONNs in the decoder at second stage models further enhances nonlinear representation, allowing the model to better capture subtle boundaries and structural variability. While the pipeline involves multiple models and incurs a higher computational cost compared to single-pass architectures, and currently includes some manual pre-processing steps, this design allows for more granular control, interpretability, and adaptability to clinical scenarios. Thus, the method provides a scalable foundation for high-accuracy, organ-specific segmentation in MRE, addressing limitations often overlooked in prior studies.

## 5. Conclusion

This study presents a novel, automated, two-stage deep learning framework for the automated segmentation of ten anatomically distinct gastrointestinal organs from MRE images. The proposed pipeline effectively integrates a coarse-to-fine strategy, combining a multiclass DenseNet201–UNet++ model for global localization with class-wise DenseNet121–SelfONN U-Net models for organ-specific refinement. This hierarchical approach enables anatomically consistent and high-precision segmentation, particularly for underrepresented and morphologically complex structures such as the appendix, cecum, and sigmoid colon. Comprehensive experiments demonstrate that the second-stage refinement significantly enhances performance across all organs, achieving substantial gains in region overlap and boundary accuracy compared to baseline models. Notably, the framework addresses key challenges in GI organs segmentation such as severe class imbalance, variable morphology, and low tissue contrast. The inclusion of SelfONN-based decoding at second stage further improves nonlinear adaptability and segmentation quality, particularly in ambiguous anatomical regions. While the current implementation is computationally intensive and relies on 2D design with a single-center dataset, these limitations highlight the importance of optimizing model efficiency and expanding data diversity to enable broader clinical adoption. Moreover, integrating explainability tools and uncertainty quantification could facilitate clinician trust and decision-making in real-world settings. The proposed method has strong potential to support MRE-based diagnostic workflows by automating organ-level analysis, reducing radiologist burden, and improving consistency in gastrointestinal disease assessment. Nonetheless, it establishes a strong foundation for future developments in organ-aware segmentation, clinical diagnostic support, and MRE-based gastrointestinal disease assessment. With continued efforts in optimization, dataset expansion, and clinical integration, this pipeline holds considerable potential for advancing automated abdominal imaging analysis and improving diagnostic workflows in gastroenterology.

## 6. Limitations and Future Directions

Despite the effectiveness of the proposed two-stage segmentation framework, several limitations should be acknowledged. First, the use of 2D coronal slices for both training and inference limits the model's ability to leverage full volumetric context, which could benefit the segmentation of spatially extended or curved organs such as the small intestine or transverse colon. Second, architecture relies on eleven independently trained models, a single coarse multiclass model followed by ten organ-specific binary models, resulting in increased computational cost for both training and deployment. This modular design, while accurate, may pose challenges for real-time or resource-constrained clinical use. Third, the dataset was derived from a single-center Crohn's disease cohort, limiting anatomical diversity and reducing generalizability to other gastrointestinal conditions or scanner settings. Lastly, although SelfONN-based decoders improved spatial precision and nonlinear representation, they introduce greater computational overhead compared to traditional convolutional architectures.

Future work should focus on integrating volumetric context through 3D or hybrid 2.5D architectures that can capture inter-slice dependencies while maintaining computational efficiency. Incorporating attention mechanisms or anatomical priors may further improve segmentation consistency in anatomically ambiguous regions. To reduce model size and speed up inference, compression techniques such as pruning, quantization, and knowledge distillation should be explored. Additionally, validating the framework on multi-institutional and multi-disease datasets is critical to ensure robustness across patient populations and imaging protocols. Enhancing the model input with multi-modal imaging (e.g., T1-weighted or diffusion-weighted MRE) and complementary clinical data could improve differentiation in low-contrast regions. Leveraging semi-supervised or self-supervised learning may also reduce annotation burden and extend applicability to broader anatomical targets. Finally, integrating interactive correction, uncertainty estimation, and explainability tools will be key to enabling safe and trustworthy deployment in clinical practice.

## 7. Institutional Review Board Statement

Not applicable

## 8. Informed Consent Statement

Not applicable

## 9. Data Availability Statement

The preprocessed data used in this study is available at reasonable request to the corresponding author.

## 10. Conflict of Interest

The authors declare no conflicts of interest for this study.

## References

[1] A. Kazempour and A. Kazempour, "Large Association of GI Tract Microbial Community with Immune and Nervous Systems," *Immunology of the GI Tract - Recent Advances*, Dec. 2022, doi: 10.5772/INTECHOPEN.104120.

[2] X. Yang et al., "Pathophysiologic Role of Neurotransmitters in Digestive Diseases," *Front Physiol*, vol. 12, p. 567650, Jun. 2021, doi: 10.3389/FPHYS.2021.567650/XML/NLM.

[3] K. A. Sharkey and G. M. Mawe, "The enteric nervous system," *Physiol Rev*, vol. 103, no. 2, pp. 1487–1564, Apr. 2023, doi: 10.1152/PHYSREV.00018.2022/ASSET/IMAGES/LARGE/PHYSREV.00018.2022_F010.JPEG.

[4] S. M. Collins and P. Bercik, "The Relationship Between Intestinal Microbiota and the Central Nervous System in Normal Gastrointestinal Function and Disease," *Gastroenterology*, vol. 136, no. 6, pp. 2003–2014, May 2009, doi: 10.1053/J.GASTRO.2009.01.075.

[5] I. Ogobuiro, J. Gonzales, K. R. Shumway, and F. Tuma, "Physiology, Gastrointestinal," *StatPearls*, Apr. 2023, Accessed: Jul. 03, 2025. [Online]. Available: https://www.ncbi.nlm.nih.gov/books/NBK537103/

[6] "illustration of Healthcare and Medical education drawing chart of Human Digestive System for Science Biology study 2803159 Vector Art at Vecteezy." Accessed: Jul. 04, 2025. [Online]. Available: https://www.vecteezy.com/vector-art/2803159-illustration-of-healthcare-and-medical-education-drawing-chart-of-human-digestive-system-for-science-biology-study

[7] Ø. Hovde and B. A. Moum, "Epidemiology and clinical course of Crohn's disease: Results from observational studies," *World J Gastroenterol*, vol. 18, no. 15, pp. 1723–1731, 2012, doi: 10.3748/WJG.V18.I15.1723.

[8] S. M. Hong and D. H. Baek, "Diagnostic Procedures for Inflammatory Bowel Disease: Laboratory, Endoscopy, Pathology, Imaging, and Beyond," *Diagnostics 2024, Vol. 14, Page 1384*, vol. 14, no. 13, p. 1384, Jun. 2024, doi: 10.3390/DIAGNOSTICS14131384.

[9] A. N. Desmond et al., "Crohn's disease: factors associated with exposure to high levels of diagnostic radiation," *Gut*, vol. 57, no. 11, pp. 1524–1529, Nov. 2008, doi: 10.1136/GUT.2008.151415.

[10] N. Shaban et al., "Imaging in inflammatory bowel disease: current and future perspectives," *Frontline Gastroenterol*, vol. 13, no. e1, pp. e28–e34, Aug. 2022, doi: 10.1136/FLGASTRO-2022-102117.

[11] J. Rimola et al., "Magnetic resonance for assessment of disease activity and severity in ileocolonic Crohn's disease," *Gut*, vol. 58, no. 8, pp. 1113–1120, Aug. 2009, doi: 10.1136/GUT.2008.167957.

[12] S. Samuel et al., "Endoscopic Skipping of the Distal Terminal Ileum in Crohn's Disease Can Lead to Negative Results From Ileocolonoscopy," *Clinical Gastroenterology and Hepatology*, vol. 10, no. 11, pp. 1253–1259, Nov. 2012, doi: 10.1016/J.CGH.2012.03.026.

[13] M. A. Khan et al., "Gastrointestinal diseases segmentation and classification based on duo-deep architectures," *Pattern Recognit Lett*, vol. 131, pp. 193–204, Mar. 2020, doi: 10.1016/J.PATREC.2019.12.024.

[14] H. Siddiki and J. Fidler, "MR imaging of the small bowel in Crohn's disease," *Eur J Radiol*, vol. 69, no. 3, pp. 409–417, Mar. 2009, doi: 10.1016/J.EJRAD.2008.11.013.

[15] K. Horsthuis, S. Bipat, R. J. Bennink, and J. Stoker, "Inflammatory Bowel Disease Diagnosed with US, MR, Scintigraphy, and CT: Meta-analysis of Prospective Studies1," *pubs.rsna.orgK Horsthuis, S Bipat, RJ*

Bennink, J Stoker*Radiology, 2008•pubs.rsna.org*, vol. 247, no. 1, pp. 64–79, Apr. 2008, doi: 10.1148/RADIOL.2471070611.

[16] W. Zhang *et al.*, "Deep convolutional neural networks for multi-modality isointense infant brain image segmentation," *Neuroimage*, vol. 108, pp. 214–224, Mar. 2015, doi: 10.1016/J.NEUROIMAGE.2014.12.061.

[17] T. Cogan, M. Cogan, and L. Tamil, "MAPGI: Accurate identification of anatomical landmarks and diseased tissue in gastrointestinal tract using deep learning," *Comput Biol Med*, vol. 111, p. 103351, Aug. 2019, doi: 10.1016/J.COMPBIOMED.2019.103351.

[18] E. Gibson *et al.*, "Automatic Multi-Organ Segmentation on Abdominal CT with Dense V-Networks," *IEEE Trans Med Imaging*, vol. 37, no. 8, pp. 1822–1834, Aug. 2018, doi: 10.1109/TMI.2018.2806309.

[19] S. Wang, Y. Cong, H. Zhu, X. Chen, … L. Q.-I. J. of, and undefined 2020, "Multi-scale context-guided deep network for automated lesion segmentation with endoscopy images of gastrointestinal tract," *ieeexplore.ieee.orgS Wang, Y Cong, H Zhu, X Chen, L Qu, H Fan, Q Zhang, M LiuIEEE Journal of Biomedical and Health Informatics, 2020•ieeexplore.ieee.org*, Accessed: Jul. 03, 2025. [Online]. Available: https://ieeexplore.ieee.org/abstract/document/9099992/

[20] Q. Vanderbecq *et al.*, "Deep learning for automatic bowel-obstruction identification on abdominal CT," *SpringerQ Vanderbecq, M Gelard, JC Pesquet, M Wagner, L Arrive, M Zins, E ChouzenouxEuropean Radiology, 2024•Springer*, vol. 34, no. 9, pp. 5842–5853, Sep. 2024, doi: 10.1007/S00330-024-10657-Z.

[21] X. Shen *et al.*, "The application of deep learning in abdominal trauma diagnosis by CT imaging," *SpringerX Shen, Y Zhou, X Shi, S Zhang, S Ding, L Ni, X Dou, L ChenWorld Journal of Emergency Surgery, 2024•Springer*, vol. 19, no. 1, Dec. 2024, doi: 10.1186/S13017-024-00546-7.

[22] Y. Gonzalez *et al.*, "Semi-automatic sigmoid colon segmentation in CT for radiation therapy treatment planning via an iterative 2.5-D deep learning approach," *Med Image Anal*, vol. 68, Feb. 2021, doi: 10.1016/j.media.2020.101896.

[23] Y. Lamash *et al.*, "Curved planar reformatting and convolutional neural network-based segmentation of the small bowel for visualization and quantitative assessment of pediatric Crohn's disease from MRI," *Journal of Magnetic Resonance Imaging*, vol. 49, no. 6, pp. 1565–1576, Jun. 2019, doi: 10.1002/jmri.26330.

[24] L. D. van Harten, C. S. de Jonge, K. J. Beek, J. Stoker, and I. Išgum, "Untangling and segmenting the small intestine in 3D cine-MRI using deep learning," *Med Image Anal*, vol. 78, May 2022, doi: 10.1016/j.media.2022.102386.

[25] N. S. Dellschaft *et al.*, "Magnetic resonance imaging of the gastrointestinal tract shows reduced small bowel motility and altered chyme in cystic fibrosis compared to controls," *Journal of Cystic Fibrosis*, vol. 21, no. 3, pp. 502–505, May 2022, doi: 10.1016/j.jcf.2021.12.007.

[26] B. Orellana, I. Navazo, P. Brunet, E. Monclús, Á. Bendezú, and F. Azpiroz, "Automatic colon segmentation on T1-FS MR images," *Computerized Medical Imaging and Graphics*, vol. 123, Jul. 2025, doi: 10.1016/j.compmedimag.2025.102528.

[27] J. Ding, Y. Zhang, A. Amjad, J. Xu, D. Thill, and X. A. Li, "Automatic Contour Refinement for Deep Learning Auto-segmentation of Complex Organs in MRI-guided Adaptive Radiation Therapy," *Adv Radiat Oncol*, vol. 7, no. 5, Sep. 2022, doi: 10.1016/j.adro.2022.100968.

[28] O. Brem, D. Elisha, E. Konen, M. Amitai, and E. Klang, "Deep learning in magnetic resonance enterography for Crohn's disease assessment: a systematic review," Sep. 01, 2024, *Springer*. doi: 10.1007/s00261-024-04326-4.

[29] Z. Zhong et al., "A comprehensive dataset of magnetic resonance enterography images with intestinal segment annotations," *Sci Data*, vol. 12, no. 1, p. 425, Dec. 2025, doi: 10.1038/S41597-025-04760-Z;SUBJMETA=1046,1503,257,639,692,699,705;KWRD=INFLAMMATORY+BOWEL+DISEASE,SCIENTIFIC+DATA.

[30] Z. Zhou, M. M. Rahman Siddiquee, N. Tajbakhsh, and J. Liang, "UNet++: A Nested U-Net Architecture for Medical Image Segmentation," *Lecture Notes in Computer Science (including subseries Lecture Notes in Artificial Intelligence and Lecture Notes in Bioinformatics)*, vol. 11045 LNCS, pp. 3–11, Jul. 2018, doi: 10.1007/978-3-030-00889-5_1.

[31] G. Huang, Z. Liu, L. Van Der Maaten, and K. Q. Weinberger, "Densely Connected Convolutional Networks," *Proceedings - 30th IEEE Conference on Computer Vision and Pattern Recognition, CVPR 2017*, vol. 2017-January, pp. 2261–2269, Aug. 2016, doi: 10.1109/CVPR.2017.243.

[32] S. Kiranyaz, J. Malik, H. Ben Abdallah, T. Ince, A. Iosifidis, and M. Gabbouj, "Self-organized Operational Neural Networks with Generative Neurons," *Neural Networks*, vol. 140, pp. 294–308, Aug. 2021, doi: 10.1016/J.NEUNET.2021.02.028.

[33] A. Rahman et al., "Deep learning-driven segmentation of ischemic stroke lesions using multi-channel MRI," *Biomed Signal Process Control*, vol. 105, Jul. 2025, doi: 10.1016/j.bspc.2025.107676.

[34] H. Phan, K. Yamamoto, T. H. Phan, and K. Yamamoto, "Resolving Class Imbalance in Object Detection with Weighted Cross Entropy Losses," Jun. 2020, Accessed: Jul. 03, 2025. [Online]. Available: https://arxiv.org/pdf/2006.01413

[35] S. Jadon, "A survey of loss functions for semantic segmentation," *2020 IEEE Conference on Computational Intelligence in Bioinformatics and Computational Biology, CIBCB 2020*, Oct. 2020, doi: 10.1109/CIBCB48159.2020.9277638.

[36] A. A. Taha and A. Hanbury, "Metrics for evaluating 3D medical image segmentation: Analysis, selection, and tool," *BMC Med Imaging*, vol. 15, no. 1, pp. 1–28, Aug. 2015, doi: 10.1186/S12880-015-0068-X/TABLES/5.

[37] D. Müller, I. Soto-Rey, and F. Kramer, "Towards a guideline for evaluation metrics in medical image segmentation," *BMC Res Notes*, vol. 15, no. 1, pp. 1–8, Dec. 2022, doi: 10.1186/S13104-022-06096-Y/FIGURES/2.

[38] J. Li et al., "Establishing a machine learning model based on dual-energy CT enterography to evaluate Crohn's disease activity," *Insights Imaging*, vol. 15, no. 1, Dec. 2024, doi: 10.1186/s13244-024-01703-x.